\documentclass[showpacs,prl,twocolumn,floatfix]{revtex4}
\usepackage{amsmath}
\usepackage{graphicx}
\usepackage{calc}
\usepackage{bbm}
\begin{document}
\author{Wolfgang Mauerer}\email{wolfgang.mauerer@ioip.mpg.de}
\author{Christine Silberhorn} \affiliation{Max-Planck Research Group
  for Optics, Information and Photonics, Junior Research Group IQO}
\title{Passive decoy state quantum key distribution: Closing the gap to
  perfect sources}  
 \date{\today}
\newcommand{\unit}[1]{\ensuremath{\text{#1}}}
\newcommand{\ie}{\emph{i.e.}}
\newcommand{\ket}[1]{\ensuremath{|#1\rangle}}
\newcommand{\bra}[1]{\ensuremath{\langle#1|}}
\newcommand{\ketbra}[2]{\ket{#1}\bra{#2}}
\newcommand{\eg}{\emph{e.g.}}
\renewcommand{\paragraph}[1]{}
\renewcommand{\subparagraph}[1]{}

\begin{abstract}
  We propose a quantum key distribution scheme which closely matches
  the performance of a perfect single photon source. It nearly attains
  the physical upper bound in terms of key generation rate and
  maximally achievable distance. Our scheme relies on a practical
  setup based on a parametric downconversion source and present-day,
  non-ideal photon-number detection. Arbitrary experimental
  imperfections which lead to bit errors are included. We select decoy
  states by classical post-processing. This allows to improve the
  effective signal statistics and achievable distance.
\end{abstract}
\maketitle

\subparagraph{Introduction} Quantum key distribution (QKD) allows two
parties (Alice and Bob) to communicate securely even in the presence
of an arbitrarily powerful eavesdropper (Eve) who tries to listen
undetected. To prove unconditional security, Eve must not be
restricted by any technological limitations, but must only be bounded
by the laws of quantum physics. A multitude of protocols has been
suggested in the last decades; BB84~\cite{Bennet1984} is the
best-known and usually best-performing protocol. It was shown to be
secure both in principle, \cite{Gottesman2002} and references therein,
and in the presence of experimental imperfections, \eg,
\cite{Gottesman2004}.  Unfortunately, the maximal distance and the bit
rates over which secure communications can be guaranteed are strongly
constricted if experimental imperfections are taken into account:
lossy channels, imperfect detectors with finite efficiency, dark counts
and misalignment errors, as well as non-ideal signal sources -- which
do not provide the required single photon
states~\cite{Gisin2002,Dusek2006} -- degrade the performance of the
protocol.  Decoy-state QKD, which was recently introduced 
by \cite{Hwang2003}, analyzed in \cite{Lo2005a,Wang2005a} and
adapted for practical use in \cite{Harrington2005,Wang2005}, could
mend this.  Still, coherent state implementations achieve only about
\(70\%\) of the maximal secure distance imposed by fundamental
physics.

\begin{figure}[htb]
  \centering\includegraphics[width=\linewidth]{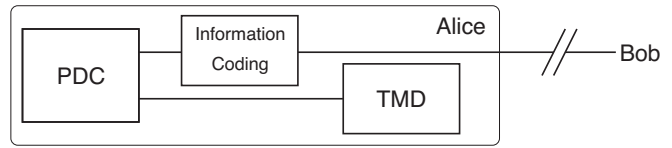}
  \caption{Setup of the proposed QKD scheme. The PDC source emits
    photon-number correlated bipartite states; the TMD records photon
    statistics.}
  \label{tmd_qkd}
\end{figure}

In this paper, we show how we can close the gap between practical QKD
implemented with state-of-the-art devices and idealized QKD assuming
perfect single-photon signals. In our approach, a parametric
downconversion (PDC) source~\cite{Mandel1995} in conjunction with a
photon number resolving detector \cite{Achilles2003} -- as depicted in
Fig.~\ref{tmd_qkd} -- is utilized to implement a passive decoy-state
QKD scheme.  It does not require any active intensity modulation, but
allows to improve the effectively sent signal statistics by employing
conjugate PDC modes. Strict photon-number correlations between the two
PDC outputs allow to infer the complete statistical information about
one of them by measuring the photon number distribution of the other.
Passive data analysis enables us to generate optimized effective
signal statistics without physical blocking. For all practical
purposes, our protocol accomplishes up to few percent the power of a
single photon source in terms of distance, while the key generation
rate is on par with the best available schemes.

\paragraph{Review of decoy QKD}
Since our work is based on the decoy state method, we begin by briefly
reviewing the basic idea.  The security of BB84  relies on single
photons, so signals with more than one photon are
insecure because Eve can perform a photon-number
splitting (PNS) attack, which has been shown to be optimal 
\cite{Lutkenhaus2002}. For this, Eve performs a quantum non-demolition
measurement of the photon number, taps one photon and delays the
measurement until Alice and Bob announce the bases. If Eve replaces
the lossy channel with a perfect one and passes on the same
statistics as in the lossy channel, this attack can not be detected in
a standard BB84 scheme.
The probability that at least one photon of an \(n\)-photon signal
passes a quantum channel with transmission \(\eta =
\exp(\alpha/10\cdot l)\) (where \(\alpha\) is the loss in
\(\unit{dB/km}\) and \(l\) is the channel length in \(\unit{km}\)) is
given by \(\eta_{n} = 1-(1-\eta)^{n}\), different loss characteristics
arise for signals with different photon numbers. The core idea of the
decoy method is to check that the signal losses behave as expected for
different photon numbers to exclude PNS attacks.  For this, it is
necessary to intersperse the signal stream with decoy states whose
intensity differs slightly from the signal states, but share all other
characteristics like wavelength and timing.  A more detailed
description can be found in the work of Lo and
coworkers~\cite{Lo2005a}.

The security analysis in~\cite{Lo2005a} proofs that a lower bound
on the secure key generation rate is given by
\begin{align} 
  S' &=  q\{-Q_{\chi}f(E_{\chi})H_2(E_{\chi}) + 
    Q_1\cdot(1-H_2(e_1))\},\nonumber\\
  S &\geq S'\cdot\Theta(S').\label{bb84_rate_locc}
\end{align}
In Eqn~\ref{bb84_rate_locc} the \emph{gain} \(Q_{\chi}\) denotes the
ratio of Bob's detection events to Alice's number of submitted signals
after sifting; the \emph{yield} \(Y_{n}\) is defined as the
probability that Bob receives a signal conditioned on that Alice has
sent an \(n\)-photon signal. The parameters \(E_{\chi}\) and \(e_{n}\)
describe the overall, and the photon number resolved quantum bit error
rate (QBER), \ie, the fraction of signals which contribute false key
bits although a signal was received. The quantities are related as
follows:
\begin{align}
  Q_{\chi} &\equiv \sum_{n=0}^{\infty}Q_{n} = 
                       \sum_{n=0}^{\infty}Y_{n}p(n),\label{sle1}\\
  E_{\chi}Q_{\chi} &\equiv \sum_{n=0}^{\infty}Y_{n}p(n)e_{n}.\label{sle2}
\end{align}
The function \(f(x)\) in Eqn.~\ref{bb84_rate_locc} accounts for
non-ideal practical error correction which does not reach the Shannon
limit, and \(H_{2}(x)\) is the binary Shannon entropy. The sifting
factor \(q\) corrects incompatible bases, \ie, for standard BB84 \(q =
1/2\).  In the asymptotic limit it is possible to reach values of \(q
\approx 1\) \cite{Dusek2006}, this is used in the remainder of the
paper.
Conventional QKD schemes employ binary detectors. Thus, only the overall
gain \(Q_{\chi}\) and QBER \(E_{\chi}\) can be measured during
transmission. Source characterization guarantees that \(p(n)\) is
known. The decoy state idea exploits that the linear system of
Equations (\ref{sle1}) and (\ref{sle2}) can be solved for \(Y_{n}\)
and \(e_{n}\), if states with different mean intensities are employed.
While \(Y_{n}\) and \(e_{n}\) are identical in Eve's absence for the
signal and all decoy states, it is proven that any PNS attack will
modify these quantities, \ie, Eve's attempt of an PNS attack will be
detected\cite{Lo2005a}.

The original security proof for BB84 given in \cite{Shor2000} utilized
local operations and one-way classical communication (1-LOCC). While
many security analyses retain with 1-LOCC, enhanced security proofs
employing 2-LOCC~\cite{Gottesman2002} have been elaborated recently
and adapted to the decoy method in~\cite{Ma2006}. Two-way
postprocessing is performed by comparing parities for random bit pairs
in Alice's and Bob's key. If the parities match, they keep the first
bit, otherwise they discard both.  One round of this procedure is
called a B-step; repeating it for several rounds is possible and
allows to increase the maximal secure distance.  For comparison we
consider both cases, 1-LOCC and 2-LOCC.

\subparagraph{Setup} Consider the setup in Fig.~\ref{tmd_qkd}.  In the
source, we use a standard PDC process to obtain the photon-number
correlated state
\begin{equation}
\ket{\psi} = \frac{1}{\mathcal{N}}\sum_{n=0}^{\infty}
\lambda_{n}\ket{n, n},\label{source_state}
\end{equation}
where \(\lambda\) and the normalization factor \(\mathcal{N}\) depend on the
physical boundary conditions~\cite{Perina2003}. The distribution
exhibits Poissonian (\(\lambda_{n} = \frac{\lambda^n}{\sqrt{n!}}\),
\(\mathcal{N} = e^{-\lambda^2}\)~\footnote{\(\chi\) is a coupling strength and
  interaction time parameter; the mean photon number is given by
  \(\sinh^{2}\chi\)~\cite{Mandel1995}.}) or thermal (\(\lambda_{n} =
\tanh^{2n}\chi\), \(\mathcal{N} = \cosh^{2}\chi\)) statistics in the extremal
cases, so we will consider both possibilities. 
Since Eve has no phase reference,
the phase is assumed to be totally randomized, and an effective
mixture of photon number eigenstates with density operator \(\varrho =
\sum_{n}|\lambda_{n}/N|^{2}\varrho_{n}\) is transmitted.

Information encoding can be accomplished by polarization or time coding,
but the exact method is of no relevance for the further analysis.
A time multiplexed detector (TMD) provides photon number resolution
capabilities. There are several methods to perform photon number
resolution, but we focus on TMD detection
\cite{Achilles2003} since it is cost-effective and easy
to handle experimentally.  The measured TMD statistics can be related
to the impinging photon number statistics by
\begin{equation}
\vec{p}_{\text{source}} = \mathbf{L}^{-1}\cdot\mathbf{C}^{-1}\cdot
\vec{p}_{\text{meas}} \equiv \mathcal{R}(\vec{p}_{\text{meas}})%
\label{inversion}
\end{equation}
where the loss matrix \(\mathbf{L}\) accounts for photon loss in the
detection, and the convolution matrix \(\mathbf{C}\) models the effect
of a finite number of detected modes in the TMD design (for details:
see \cite{Achilles2003}); \(\vec{p}_{\text{s}} \) and \(
\vec{p}_{\text{m}}\) describe the original photon number distribution
of the source and the measured statistics.
The quantity \(\mathbf{C}\cdot\mathbf{L}\) can be determined by
measurement, but there is also an analytical representation
\(p_{\eta}(m|n)\) for the matrix entries given in \cite{Fitch2003}.
It represents the probability to get an \(m\)-photon detection outcome
conditioned on \(n\) photons entering the detector with total loss
\(\eta\). Using Eqn.~\ref{inversion}, the TMD measurement can be
inverted so that the real statistics of the source are reconstructed
with high fidelity\cite{Achilles2006}.  Note that this inversion is
only possible for an ensemble of states, but not for a single signal,
so Alice needs to record the measurement results of the TMD for every
time slot.  After this, Alice and Bob follow the standard protocol of
BB84~\cite{Gisin2002} for information encoding and analysis.

\begin{figure}[htb]
  \centering\includegraphics[width=\linewidth]{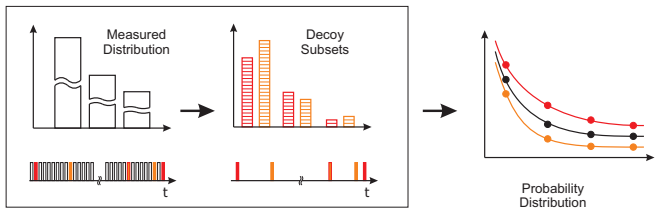}
  \caption{(Color online) Classical decoy state selection. Apt subsets
    of the recorded TMD measurements are selected and inverted to form
    the decoy states which are similar to the signal state. The photon
    number distributions are not drawn to scale for instructive
    purposes.}\label{decoy_sel}
\end{figure}

\subparagraph{Decoy generation} The essential step of passive decoy
state selection follows after a sufficiently large number of signals
(say, \(N_\text{tot} \gg 1\)) has been transmitted; note that Alice
will run the source with constant pump intensity and without any
active optical manipulations for the duration of the procedure.
Fig.~\ref{decoy_sel} provides an overview about the process: The
measured discrete probability distribution \(\vec{p}_\text{meas}\) is
calculated by \(p_{\text{meas}}(n) =
\frac{\#n_{\text{tot}}}{N_{\text{tot}}}\), where \(\#n_{\text{tot}}\)
denotes the number of \(n\)-photon measurement outcomes from the TMD.
This distribution can be inverted by Eqn.~\ref{inversion}; the strict
photon number correlations of the PDC states ensure that Alice's
measurement coincide with the signal statistics.

Assume that we start with a random selection of a set containing \(M
\ll N_{\text{tot}}\) signals to construct decoy states, which have
exactly the same statistics as the remaining signal states.  Alice
then additionally picks \(\delta_{n}\) slots with an
\(n\)-photon measurement result such that

\begin{equation}
  \#n_{\text{decoy}} = \#n_{\text{tot}}\cdot \frac{M}{N_{\text{tot}}} 
                     + \delta_{n}
\end{equation}
where \(\delta_{n}\) is a small positive or negative offset which
results in a photon number distribution of the decoy subset differing
slightly from the distribution in the signal. The decoy subset can be
inverted to obtain the proper probability distribution
\(\vec{p}_{\text{decoy}}\).  Depending on how many decoy states are to
be used (one vacuum state and two decoy states similar to the signal
are sufficient, cf.~\cite{Ma2005,Harrington2005}), an appropriate
number of subsets has to be chosen. Different distributions of
\(\delta_{n}\) for different subsets ensure that the generated decoy
signals are sufficiently distinct from each other as required to solve
the system of linear equations~(\ref{sle1},\ref{sle2}).  We would like
to stress that our passive method for ``generating'' decoys provides
distinct advantages: during signal transmission it is still undecided 
which states will become signal or decoy states.  This implies that a
distinction between signal and decoy states by Eve is not possible,
even in principle. It is also important to emphasize that
our decoy selection mechanism eliminates many experimental challenges
(\eg, different spectra of the generated PDC light for signal and
decoys which gives Eve a chance to experimentally distinguish between
them) which arise in proposals with the same hardware, but a different
analysis procedure (\cite{Horikiri2006, Cai2006}), which do not draw
maximal use of the TMD's capabilities. 
%
The remainder of the protocol is identical to a standard decoy scheme:
Alice and Bob check \(e_{n}\) and \(Y_{n}\) as described above.
Error correction and privacy amplification need to be performed to
generate a final secure key. The inset in Fig.~\ref{rate_simulation}
presents our simulation results (for details see below). The key
generation rate and maximal secure distance closely match a perfect
single photon source.

\subparagraph{Statistics enhancement} 
The TMD results can not only be used to generate decoy states, but
also provide improved effective signal statistics.  
While the error rates \(e_{n}\) for \(n \geq 1\) are the order of
\(10^{-2}\), the contribution by vacuum signals is \(e_{0} =
1/2\)~\footnote{If a vacuum pulse is sent and a dark count causes one
  detector to click, it is the wrong one with \(50\%\) chance.}.
Thus, it is desirable to remove such events as good as 
possible. 
Decreasing the dark count rate on Bob's side is hard because it
requires refinement of the detectors, while fine-grained time
triggering can be used on Alice's side to reduce the dark count
probability in the TMD to a negligible level, \ie, \(p(n|m) = 0\) for
\(n > m\) \cite{Achilles2003}.
Note that due to losses and imperfect detection, filtering
multi-photon contributions does not work perfectly and results in
comparatively small rate improvements (\cite{Horikiri2006}).  The
benefits are negligible in contrast to filtering zero photon
contributions.
Alice has recorded the TMD measurement for every signal. Hence she can
easily discard all zero events in the postprocessing phase which
leads to a better effective probability distribution given by
\begin{equation}
  p_{f,\text{meas}}(n) = \begin{cases}
    0 & n = 0\\
    \frac{1}{N_{\text{tot}}-\sum_{n = 1}^{\infty}\#n}\cdot
                            \frac{\#n}{N_{\text{tot}}}  & n \geq 0
  \end{cases},\label{filter}
\end{equation}
where \(p_{f,\text{meas}}\) denotes the measured, filtered
distribution; the effectively sent distribution is \(\vec{p} =
\mathcal{R}(\vec{p}_{f,\text{meas}})\).  Since \(p(0|n) \neq 0\) for
\(n > 0\), some usable signal states are also removed from the
distribution, but this does not endanger the total positive effect of
the filtering.
To implement the operation, Alice and Bob need to discard all slots in
the postprocessing stage where the TMD result was zero and use the
inverted probability distribution in the rate calculations.

\begin{figure}[htb]
  \centering\includegraphics[width=\linewidth]{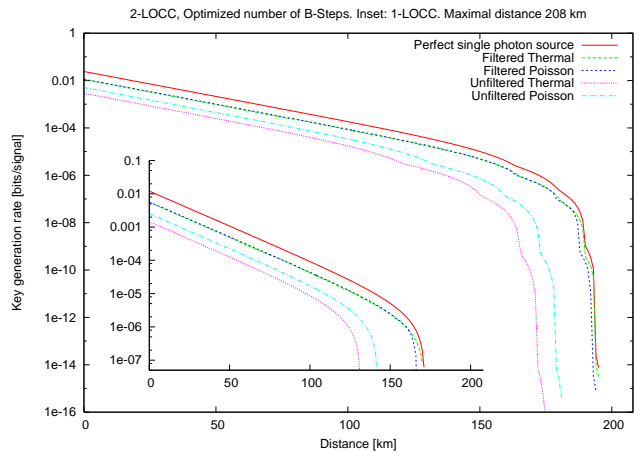}
  \caption{(Color online) Simulation results for two-way and one-way
    (inset) classical communication. Both graphs were obtained by a
    numerical evaluation of Eqn.~\ref{bb84_rate_locc}; the optimal
    values for \(\chi\) and the number of B-Steps which maximize the
    key generation rate have been used for all distances. The right
    border represents the principal upper bound given by the
    intercept-resend attack.}
  \label{rate_simulation}
\end{figure}

Note that filtering of this type does not reduce the signal
rate: A subset of the signal set is removed during postprocessing
because this subset will make the overall result only
worse. No physical blocking of signals is performed.  This
leaves the transmitted signals unmodified.
\subparagraph{Numerical simulations} 
Fig.~\ref{rate_simulation} and Tab.~\ref{dist_comp} present the
results of the numerical evaluation for all cases discussed above:
Signals with and without filtering for both 1- and 2-LOCC. Note that
we apply an optimization for both, the best value for \(\chi\) and the
ideal number of B-Steps for every distance. To allow comparison with
other proposals, we use the set of experimental parameters given in
\cite{Gobby2004}. The upper bound on the secure distance caused by the
undetectable intercept-resend attack at a QBER of more than
\(25\%\)~\cite{Dusek2006} lies at \(208 \unit{km}\), \ie, the right
border of the graph.  The \emph{lower} bound on our rate closely
approaches this \emph{upper} limit, and reaches the single photon
performance.  One also needs to keep in mind that this upper bound is
not even tight, but can be replaced by smaller ones (\eg,
\cite{Moroder2006}).



The filtering transformation in Eqn.~\ref{filter} modifies the
effective signal distribution so that a different rate is obtained although
the sent statistics remain unmodified. Thus, a penalty factor needs to
be introduced into Eqn.~\ref{bb84_rate_locc} when an optimal
\(\chi\) is sought to maximize \(S\):
%
\(
S \geq p_{\text{pen}}\cdot S'\cdot \Theta(S'), \hspace{2mm}
p_{\text{pen}} =  1-\sum_{n=0}^{\infty} p(0|n)p_{s}(n).
\)
%
%
The optimal values for \(\chi\) depend on the simulation parameters
and the source statistics; a comprehensive set of results for
different combinations can be found in \cite{Mauerer}. Here it
suffices to know that the range for \(\chi\) is \((0,0.5)\) which can
well be realized with current PDC sources \cite{U'Ren2004}. 
Existing sources provide better performance than actually required.

\begin{table}[htb]
  \begin{tabular}{lc@{\hspace*{5mm}}c@{\hspace*{2mm}}c}
    Source & \hbox{\vbox{\hbox{distance}\hbox{unf./filt.}}} &
    \(\Delta_{1,\text{f}}\) &
    \(\Delta_{2,\text{f}}\)
    \\\hline\hline
    Thermal (1-way)        & 130.8/169.7 & 0.7\% & 18.3\% \\
    Thermal (2-way)        & 174.5/194.5 & 0.4\% & 6.3\% \\\hline
    Poissonian (1-way)     & 141.2/166.0 & 2.9\% & 20.0\% \\
    Poissonian (2-way)     & 180.8/193.8 & 0.7\% & 6.6\%  \\
  \end{tabular}
  \caption{Comparison of the obtainable distances for different signal
    sources and postprocessing methods with the limits set by a perfect single
    photon source and the principal physical upper bound.
    A perfect single photon source achieves \(170.9 \unit{km}\) for
    1-LOCC and \(195.2 \unit{km}\) for 2-LOCC.
    \(\Delta_{1,\text{f}}\) denotes the difference to this distance.
    \(\Delta_{2,\text{f}}\) denotes the the
    difference to the principal intercept-resend upper bound.
    Both refer to the effectively filtered source.
    At most 4 B-Steps were used.}\label{dist_comp}
\end{table}

As explained above, two-way processing with B-steps can increase the
achievable distances. Ma \emph{et al.}~\cite{Ma2006} calculated that
after performing a B-Step, a lower bound on the secure key generation
rate is given by
%
\(  S' = q Q_{\chi}\left(\frac{1}{2}s_{n\neq 1}\left(
      -f(E_{\chi}')H_{2}(E_{\chi}') + 
      \Omega'(1-H_{2}(e_{1,p}'))
    \right)\right)\),
  \(S \geq S'\cdot\Theta(S')\).
%
A detailled derivation of the formula is beyond the scope of this paper, but 
can be found in Refs.~\cite{Ma2006,Mauerer}.
Multiple rounds of B-Steps apply the transformation multiple times).
%
%
%
The difference between the lower bound on the maximal secure
distance and the principal limit shrinks to about \(6.5\%\) with 4
B-Steps as shown in Tab.~\ref{dist_comp}.

\subparagraph{Conclusions} 
In summary, we have shown how to use the
photon number correlations of a PDC source to implement a BB84 scheme
which nearly reaches the performance of a single photon scheme. This removes
the predominant imperfection from real-world QKD implementations.
Since the lower bound on the key generation rate coincides up to a few
percent with the principal upper bounds, further improvements need
either come from new protocols or improved hardware. Refinements of
security proofs will likely be unfruitful by comparison.

We acknowledge helpful comments by N.~L\"utkenhaus, H.-K.~Lo and
J.~Lundeen.
\bibliography{literature}
\end{document}